\documentstyle[12pt,epsfig]{article}

\def\be{\begin{equation}}
\def\ee{\end{equation}}
\def\la{\langle}
\def\ra{\rangle}
\def\IP{\hbox{\rm I\kern -1.6pt{\rm P}}}
\def\IC{{\hbox{\rm C\kern-.58em{\raise.53ex\hbox{$\scriptscriptstyle|$}}
    \kern-.55em{\raise.53ex\hbox{$\scriptscriptstyle|$}} }}}
\def\IN{\hbox{I\kern-.2em\hbox{N}}}
\def\IR{\hbox{\rm I\kern-.2em\hbox{\rm R}}}
\def\ZZ{\hbox{{\rm Z}\kern-.3em{\rm Z}}}
\def\IT{\hbox{\rm T\kern-.38em{\raise.415ex\hbox{$\scriptstyle|$}} }}

\newtheorem{theorem}{Theorem}[section]

\newtheorem{proposition}[theorem]{Proposition}
\newtheorem{corollary}[theorem]{Corollary}

\begin{document}

\title{On a slow drift of a massive piston in an ideal gas
that remains at mechanical equilibrium}
\author{N. Chernov\\ Department of Mathematics\\
University of Alabama at Birmingham\\
Birmingham, AL 35294, USA\\
chernov@math.uab.edu}
\date{\today}
\maketitle

\begin{abstract}
We consider a heavy piston in an infinite cylinder surrounded by
ideal gases on both sides. The piston moves under elastic
collisions with gas atoms. We assume here that the gases always
exert equal pressures on the piston, hence the piston remains at
the so called mechanical equilibrium. However, the temperatures
and densities of the gases may differ across the piston. In that
case some earlier studies by Gruber, Piasecki and others reveal a
very slow motion (drift) of the piston in the direction of the
hotter gas. At the same time the hotter gas slowly transfers its
energy (heat) across the piston to the cooler gas. While the
previous studies of this interesting phenomenon were only
heuristic or experimental, we provide first rigorous proofs
assuming that the velocity distribution of the ideal gas satisfies
a certain ``cutoff'' condition.

\end{abstract}

\renewcommand{\theequation}{\arabic{section}.\arabic{equation}}

\section{Introduction}
\label{secI} \setcounter{equation}{0}

Consider an isolated cylinder filled with an ideal gas and divided into
two compartments by a large piston which is free to move along the axis
of the cylinder, Fig.~1. The piston interacts with the gas atoms via
elastic collisions. Assume that the gas in each compartment separately
is at equilibrium with temperature and density $T_-,n_-$ and $T_+,n_+$,
respectively. Let the gases exert equal pressures on the piston, i.e.\
let
$$
           P_- = n_-k_BT_- = n_+k_BT_+ = P_+
$$
Then the system is at the so called {\em mechanical equilibrium}, and
according to the laws of thermodynamics this state should be
(macroscopically) stable.

\begin{figure}[h]
\centering
\epsfig{figure=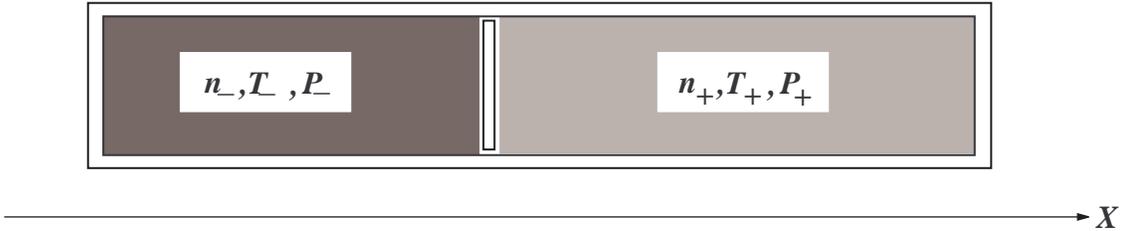}\caption{Piston in a cylinder filled with
gas.}
\end{figure}

However, the system as a whole is not in a true equilibrium state,
unless $T_-=T_+$, hence microscopically it is not stable yet and should
find ways to evolve to a true, {\em thermal} equilibrium, in which $T_-
= T_+$. This was predicted earlier by Landau and Lifshitz \cite{LL},
Feynman \cite{F} and others. Recently Gruber, Piasecki and Frachebourg
and others \cite{GP,GF,P} derived heuristically, by means of kinetic
theory, exact formulas describing the slow drift of the piston and the
slow heat transfer from the hotter gas to the cooler gas.

Our goal is to derive rigorously the main formulas of \cite{GP,GF,P}
describing the slow drift of the piston and the slow heat transfer
between the gases.

The piston model trivially reduces to a one-dimensional system by the
projection onto the axis of the cylinder. Then one obtains an ideal gas
on an interval, and the piston itself becomes a heavy point particle.
The motion of a heavy particle in an infinite ideal gas of light
particles is a classical example of Brownian motion studied by van
Kampen \cite{vK} and many others \cite{L,H,DGL,GF,GP,P}.

We consider a one-dimensional ideal gas on the entire line, without
boundaries. The heavy piston is initially placed at the origin $X(0)=0$
and is at rest $V(0)=0$. The initial configuration of gas atoms and
their velocities is chosen at random as a realization of a
(two-dimensional) Poisson process on the $(x,v)$-plane with density
$p(x,v)$. This means that for any domain $D\subset \IR^2$ the number
$N_D$ of gas particles $(x,v)\in D$ at time $t=0$ is a Poisson random
variable with parameter
$$
          \lambda_D=\int\!\!\int_D p(x,v)\, dx\, dv
$$
The system evolves according to the rules of elastic collisions. Denote
the mass of the piston by $M$ and the mass of an atom by $m$. Since
atoms have identical masses, their collisions can be ignored. When an
atom with velocity $v$ collides with the piston, whose velocity is $V$,
their velocities after the collision are given by
\be
    V^\prime = \frac{M-m}{M+m}\, V + \frac{2m}{M+m}\, v
      \label{V'}
\ee
\be
    v^\prime = -\frac{M-m}{M+m}\, v + \frac{2M}{M+m}\, V
      \label{v'}
\ee
These rules preserve the total kinetic energy and the total momentum.
Between collisions, all the particles and the piston move with constant
velocities.

The position $X(t)$ and velocity $V(t)=dX(t)/dt$ of the piston make a
random process whose characteristics are determined by the initial gas
density $p(x,v)$. It is natural to assume that $p(x,v)$ is symmetric in
$v$ and spatially homogeneous, i.e. $p(x,v)=p(v)$ and $p(v)=p(-v)$. We
can set $p(v)=nf(v)$, where $f(v)=f(-v)$ is a probability density and
$n>0$ is a (constant) spatial density. Then one can approximate $X(t)$
and $V(t)$ by certain Gaussian stochastic processes:

\begin{theorem}[Holley \cite{H}]
Let the density $f(v)$ have a finite fourth moment $\int v^4f(v)\,
dv<\infty$. Then for every finite $t_0<\infty$, the function
$V(t)\sqrt{M}$ on the interval $[0,t_0]$ converges, in distribution, as
$M,n\to\infty$ and $M/n\to\,{\rm const}$, to an Ornstein-Uhlenbeck
velocity process ${\cal V}_t$, while $X(t)\sqrt{M}$ converges to an
Ornstein-Uhlenbeck position process ${\cal X}_t$. \label{tmH}
\end{theorem}

An Ornstein-Uhlenbeck process $({\cal X}_t,{\cal V}_t)$ is defined by
\cite{Ne}
$$
    d{\cal X}_t = {\cal V}_t\, dt,
    \ \ \ \ \ \
    d{\cal V}_t = -a{\cal V}_t\, dt +\sqrt{D}\, d{\cal W}_t
$$
where $a>0$, $D>0$ are constants and ${\cal W}_t$ a Wiener
process. The Ornstein-Uhlenbeck position process ${\cal X}_t$
converges in an appropriate limit (e.g.\ $a\to\infty$,
$a^2/D=\,$const) to a Wiener process.

We note that D\"urr et al \cite{DGL} extended the above theorem to
arbitrary dimensions.

Our paper concerns with another physically interesting situation, where
the initial density is {\em not} spatially homogeneous, but a spatial
homogeneity is assumed separately for the gases to the right and to the
left of the piston. So we assume that
$$
     p(x,v)=\left\{\begin{array}{cc}
     p_+(v) & {\rm for}\ x>0\\
     p_-(v) & {\rm for}\ x<0
     \end{array}\right .
$$
and we also assume symmetry: $p_-(v)=p_-(-v)$ and $p_+(v)=p_+(-v)$. We
can set $p_{\pm}(v)=n_{\pm}f_{\pm}(v)$, where $f_{\pm}(v)=f_{\pm}(-v)$
are probability densities and $n_{\pm}>0$ are (constant) spatial
densities.

In addition, we want to exclude macroscopic motion of the piston in
either direction. The macroscopic velocity of the piston vanishes if
and only if
\be
    n_- \int_0^{\infty} v^2f_-(v)\, dv =
    n_+ \int_0^{\infty} v^2f_+(v)\, dv
        \label{V=0}
\ee
as it was shown heuristically in \cite{LPS} and under some conditions
rigorously in \cite{CLS}. The physical interpretation of Eq.\
(\ref{V=0}) is the pressure balance. If we define the pressures of the
gases by
$$
        P_{\pm} = m n_{\pm}
        \int_{-\infty}^{\infty} v^2 f_{\pm}(v)\, dv
$$
(which would be a proper thermodynamical pressure if the velocity
distributions were Maxwellian), then the condition (\ref{V=0}) means
exactly that $P_-=P_+$.

To simplify some technical considerations we assume that the initial
densities of the gases satisfy the following {\em velocity cutoff}:
\be
     p_{\pm}(v) = 0, \quad {\rm if} \quad |v| \leq v_{\rm min}
     \quad {\rm or} \quad |v| \geq v_{\max}
        \label{cutoff}
\ee
for some $0<v_{\min}<v_{\max}<\infty$. Hence, the initial velocities of
atoms are bounded away from zero and infinity. Under these conditions,
our arguments are rigorous. We also discuss in Section~\ref{secJMA} how
to relax these assumptions, leaving this work for the future.

\section{Markov approximation}
\label{secMA} \setcounter{equation}{0}

Our cutoff condition (\ref{cutoff}) has an important implication. As
long as the speed of the piston remains small enough, every gas atom
collides with the piston at most once. Indeed, let $|V(t)|< V_{\max}$,
where
$$
     V_{\max} = \frac{M-m}{3M+m}\, v_{\min}
$$
Note that $V_{\max}$ is close to $v_{\min}/3$ when $M\gg m$. Then
it follows directly from (\ref{cutoff}) and (\ref{v'}) that the
atoms' velocities after collisions are at least
$$
    \frac{M-m}{M+m}\, v_{\min}-\frac{2M}{M+m}\, V_{\max}
      = V_{\max}
$$
the last equality holds due to our choice of $V_{\max}$. Therefore, the
atoms after collisions remain faster than the piston, so the latter
cannot ``catch up'' with them.

Hence, as long as $|V(t)|<V_{\max}$, the velocity of the piston $V(t)$
evolves as a Markov process with piecewise constant trajectories (a
jump or step process). Moreover, this is a stationary (homogeneous in
time) Markov process due to our requirements on the density $p(x,v)$.
We will slightly change the function $V(t)$ so that it will evolve as a
stationary Markov process unconditionally. Fix some $\bar{V}\in
(0,V_{\max})$ and require that whenever the velocity $V'$ of the piston
after a collision, in the notation of (\ref{V'}), exceeds $\bar{V}$ in
absolute value, it gets ``reflected'' at $\bar{V}$, i.e.\ it
instantaneously changes from $V'$ to $V''$ by the rules
\be
   \begin{array}{lcl}
   V'>+\bar{V} & \Longrightarrow & V''=+2\bar{V}-V'\\
   V'<-\bar{V} & \Longrightarrow & V''=-2\bar{V}-V'
   \end{array}
\ee
These rules are in the spirit of random walks with reflecting
boundary conditions.

We will denote the velocity of the piston in the so defined dynamics by
$W(t)$. It is clear that $W(t)=V(t)$ for all $t<T$ such that
$\sup_{0<t<T}|V(t)|<\bar{V}$. We first study the Markov process $W(t)$
and later estimate the difference $V(t)-W(t)$.

Denote by ${\cal P}(u,d w;\Delta t)$ for every $\Delta t>0$ the
transition probability for the process $W(t)$, i.e.\
$$
    P\left (\, W(t+\Delta t)\in A/W(t)=u\, \right )
     = \int_A {\cal P}(u,d w;\Delta t)
$$
for every Borel set $A\subset\IR$. It is clear that the piston does not
experience any collisions with the atoms during the interval
$(t,t+\Delta t)$ if and only if the trapezoidal domain
\be
           D = D(t,\Delta t)=
           \left \{(x,v):\  \frac{v-u}{x-X(t)}
           < -\frac{1}{\Delta t},\ \ v_{\min}<|v|<v_{\max}\right \}
              \label{D}
\ee
does not contain gas atoms. Therefore, the probability that
$W(t+\Delta t)=W(t)=u$ is
\be
      {\cal P}(u,\{u\};\Delta t)=\exp\left (
      -\int_{D}p(x,v)\, dx\, dv\right )
         \label{Pww}
\ee
To evaluate this integral, we partition the domain $D=D(t,\Delta
t)$ into two trapezoids, $D=D^-\cup D^+$ as shown on Fig.~2. Under
our assumptions (\ref{cutoff})
$$
        \int_D p(x,v)\, dx\, =
        \Delta t\left [ \int_{v_{\min}}^{v_{\max}}
        (v-u)\, p_-(v)\, dv +\int_{-v_{\max}}^{-v_{\min}}
        (u-v)\, p_+(v)\, dv \right ]
$$
We introduce the following notation: for each $k\geq 0$ let
$$
     \int_{v_{\min}}^{v_{\max}} v^kp_-(v)\, dv = F_k^-
      \ \ \ \ {\rm and} \ \ \ \
     \int_{-v_{\max}}^{-v_{\min}} v^kp_+(v)\, dv = F_k^+
$$
and
$$
          Q_k = F_k^- - F_k^+
$$
Then we obtain
\begin{eqnarray}
      \int_D p(x,v)\, dx\,
       &=&
        \Delta t\left [ F_1^- - F_1^+ -
        (F_0^- - F_0^+)u\right ]\nonumber\\
        &=&
        \Delta t\left [ Q_1 - Q_0 u\right ]
          \label{intDp}
\end{eqnarray}
where $u=W(t)$.

\begin{figure}[h]
\centering
\epsfig{figure=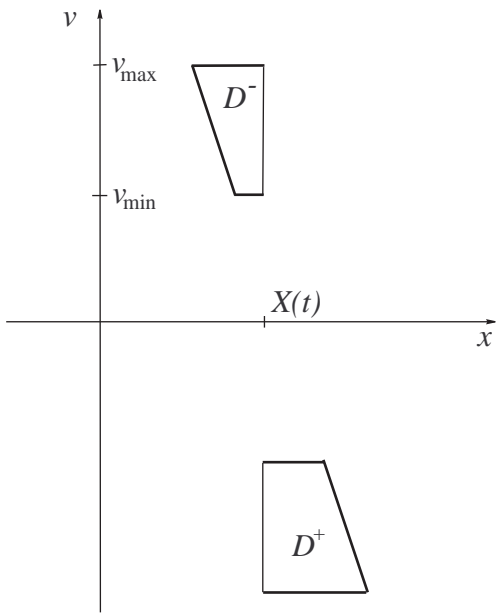} \caption{Region $D=D^-\cup D^+$.}
\end{figure}

It is clear that for every $u$ and $\Delta t>0$ the probability
measure ${\cal P}(u,d w;\Delta t)$, besides having an atom at $u$
with probability given by (\ref{Pww}), has an absolutely
continuous component with a positive density on the interval
$(-\bar{V},\bar{V})$ (in fact, that density is bounded away from
zero for every fixed $u$ and $\Delta t>0$).

\begin{proposition}
The stationary Markov process $W(t)$ has a unique stationary
measure $\mu_0$, which is absolutely continuous on the interval
$(-\bar{V},\bar{V})$. Any other initial distribution converges to
$\mu_0$ uniformly exponentially fast in time.
\end{proposition}

\noindent{\em Proof}. The process $W(t)$ satisfies the so called
Doeblin condition \cite{D}: there exist $\varepsilon>0$ and
$\Delta t>0$ such that for every Borel set $A\subset (-\bar{V},
\bar{V})$ and every $u\in (-\bar{V}, \bar{V})$ we have
$$
     m(A)<\varepsilon \Longrightarrow
     \int_A {\cal P}(u,dw;\Delta t) < 1-\varepsilon
$$
where $m(A)$ is the Lebesgue measure of $A$. Then our result
follows from general theorems \cite{D}. But we also outline a
simple direct argument.

Since the transition probability ${\cal P}(u,d w;\Delta t)$ defines,
for each $\Delta t>0$, a continuous map on the convex space of
probability distributions on the interval $(-\bar{V},\bar{V})$, the
existence follows from a general Schauder-Tychonoff theorem \cite{DS},
p.\ 456. Since ${\cal P}(u,d w;\Delta t)$ has an absolutely continuous
component whose density is bounded away from zero, the stationary
distribution $\mu_0$ is absolutely continuous with density $f_0(w)\geq
c_0>0$.

To prove the uniqueness of $\mu_0$, suppose $\mu_0'\neq\mu_0$ is
another stationary measure with density $f_0'(w)\geq c_0'>0$. Then
$(1+c)f_0-cf_0'$ for small $c>0$ is also a stationary density. Let
$$
      \bar{c}=\sup\{c>0:\ \inf_{|w|<\bar{V}}\,
      [(1+c)f_0(w)-c f_0'(w)]\geq 0\}
$$
Then $(1+\bar{c})f_0-\bar{c}f_0'$ is a stationary density, hence
it must also be bounded away from zero on $(-\bar{V},\bar{V})$.
But this contradicts our choice of $\bar{c}$.

To prove the convergence, we take an arbitrary initial
distribution $\nu_0$ and consider its image $\nu_t$ at time $t$.
Fix a $t>0$. The measure $\nu_{t}$ has an absolutely continuous
component whose density is bounded away from zero. Hence $\nu_{t}
= (1-\alpha)\nu^{(1)}+ \alpha\mu_0$ with some small $\alpha>0$ and
some measure $\nu^{(1)}$. The value of $\alpha$ depends on $t$ but
not on $\nu_0$. This implies, by induction, that
$\nu_{kt}=(1-\alpha)^k\nu^{(k)}+[1-(1-\alpha)^k]\mu_0$ for all
$k\geq 1$, which proves the exponential convergence of $\nu_t$ to
$\mu_0$. $\Box$\medskip

We now derive one useful equation. Fix a $\Delta t>0$ and for
every $u\in(-\bar{V},\bar{V})$ denote by
$$
       E(u,\Delta t) = \int w\, {\cal P}(u,dw;\Delta t)
$$
the conditional expectation of the process $W(t+\Delta t)$ given
that $W(t)=u$. The invariance of the measure $\mu_0$ implies
\be
        \int E(u,\Delta t)\, d\mu_0(u) = \int w\, d\mu_0(w)
          \label{mart}
\ee
Now suppose that
\be
           E(u,\Delta t) = u + g(u)\,\Delta t + R(u,\Delta t)
              \label{Eg}
\ee
so that
$$
       |R(u,\Delta t)|\leq\,{\rm const}\cdot (\Delta t)^2
$$
Substituting this into (\ref{mart}) and taking the limit as
$\Delta t\to 0$ gives
\be
         \int g(u)\, d\mu_0(u) = 0
            \label{g0}
\ee

Our next step is to derive (\ref{Eg}) for small $\Delta t>0$ and
compute $g(u)$ explicitly. Let $X(t)=X$ and $W(t)=u$. Due to our
assumption (\ref{cutoff}) the probability that the domain
$$
           D_1 =
           \left \{(x,v):\  0<(X-x)\cdot {\rm sgn}\, v<\bar{V}\Delta t,
           \ \ v_{\min}<|v|<v_{\max}\right \}
$$
contains more than one atom is ${\cal O}((\Delta t)^2)$, hence we can
ignore it and assume that $D_1$ contains at most one atom. Note that
$D_1$ is much larger than $D$ defined by (\ref{D}). Now it is easy to
see that the piston experiences at most one collision during the
interval $(t,t+\Delta t)$. Precisely, a collision occurs if and only if
the domain $D=D^+\cup D^-$ contains an atom. When it does, and the atom
has velocity $v$ at time $t$, then the piston's velocity at time
$t+\Delta t$ is computed by (\ref{V'}):
\be
     W(t+\Delta t)=u+\frac{2m}{M+m}\, (v-u)
       \label{WW}
\ee
Therefore, the conditional expectation $E(u,\Delta t)$ is
\be
    E(u,\Delta t)=u+\frac{2m}{M+m}\,\left [
    \int_D vp\, dx\, dv - u\int_D p\, dx\, dv\right ]
    +{\cal O}((\Delta t)^2)
       \label{EEE}
\ee
(the last term comes from our assumption that $D$ contains at most
one atom). While the second integral in (\ref{EEE}) is already
computed in (\ref{intDp}), the calculation of the first one is
similar. For every $k\geq 0$ we have
\begin{eqnarray}
        \int_D v^k\, p(x,v)\, dx\, &=&
        \Delta t\left [ \int_{v_{\min}}^{v_{\max}}
        v^k(v-u)\, p_-(v)\, dv +\int_{-v_{\max}}^{-v_{\min}}
        v^k(u-v)\, p_+(v)\, dv \right ]\nonumber\\
        &=&
        \Delta t\left [ F_{k+1}^- - F_{k+1}^+ -
        (F_k^- - F_k^+)u\right ]\nonumber\\
        &=&
        \Delta t\left [ Q_{k+1} - Q_k u\right ]
          \label{intDvp}
\end{eqnarray}
Thus,
$$
    E(u,\Delta t)=u+\frac{2m}{M+m}\,\left [
    Q_2 - 2Q_1u + Q_0u^2\right ]\Delta t
    +{\cal O}((\Delta t)^2)
$$
This gives us
$$
    g(u) = \frac{2m}{M+m}\,\left [
    Q_2 - 2Q_1u + Q_0u^2\right ]
$$
We now apply the key identity (\ref{g0}) and arrive at
$$
    Q_2 - 2Q_1\la u\ra + Q_0 \la u^2\ra = 0
$$
where $\la\cdot\ra$ means averaging with respect to the stationary
measure $\mu_0$ (we note that $Q_i$ do not depend on $u$). Thus, the
average velocity of the piston is given by
\be
      \la W\ra = \frac{Q_0\la W^2\ra + Q_2}{2Q_1}
        \label{W0}
\ee

The mechanical equilibrium is defined by the equality of pressures on
both sides of the piston, i.e.\ by
$$
         Q_2 = F_2^- - F_2^+ = 0
$$
and we will restrict ourselves to this case in the rest of the
paper. Then (\ref{W0}) reduces to
\be
      \la W\ra = \frac{Q_0\la W^2\ra}{2Q_1}
        \label{W00}
\ee
which is still different from zero, albeit very small.

It is well known that at mechanical equilibrium the average value of
$\la W^2\ra$ is ${\cal O}(m/M)$, see, e.g.\ \cite{H,DGL}. But we will
estimate it more accurately below.

Similarly to $E(u,\Delta t)$, we define
$$
       E_2(u,\Delta t) = \int w^2\, {\cal P}(u,dw;\Delta t)
$$
the conditional expectation of the square velocity $W^2(t+\Delta
t)$ given that $W(t)=u$. The stationarity of the distribution
$\mu_0$ implies
$$
        \int E_2(u,\Delta t)\, d\mu_0(u) = \int u^2\, d\mu_0(u)
$$
Assume that
$$
        E_2(u,\Delta t) = u^2 + g_2(u)\, \Delta t
        + R_2(u,\Delta t)
$$
with $|R_2(u,\Delta t)|\leq\,{\rm const}\cdot(\Delta t)^2$. Then,
as in (\ref{g0})
\be
         \int g_2(u)\, d\mu_0(u) = 0
            \label{g2}
\ee
Now, squaring the equation (\ref{WW}) we obtain
$$
    W^2(t+\Delta t)=u^2+\frac{4m}{M+m}\, (v-u)u
    + \frac{4m^2}{(M+m)^2}\, (v-u)^2
$$
Then, as in (\ref{EEE}),
\begin{eqnarray*}
  E_2(u,\Delta t) = u^2 &+&
  \frac{4m}{M+m}\int_D(v-u)up\, dx\, dv\\ &+&
  \frac{4m^2}{(M+m)^2}\int_D(v-u)^2p\, dx\, dv +
  {\cal O}((\Delta t)^2)
\end{eqnarray*}
Computing these integrals as before gives
\begin{eqnarray*}
     g_2(u) &=&
     \frac{4m}{M+m}\,(Q_2u-2Q_1u^2+Q_0u^3) \\
     & & + \frac{4m^2}{(M+m)^2}\,(Q_3-3Q_2u+3Q_1u^2-Q_0u^3)
\end{eqnarray*}
We included the terms with $Q_2$ for the sake of clarity, even
though we had assumed that $Q_2=0$. In the subsequent expressions
we remove all $Q_2$'s. The equation (\ref{g2}) gives
$$
     2Q_1\la W^2\ra - Q_0\la W^3\ra =
     \frac{m}{M+m}\,(Q_3+3Q_1\la W^2\ra - Q_0\la W^3\ra )
$$
For brevity, we denote $\varepsilon = m/M$. By making one step
further and analyzing $E_3(u,\Delta t)=\int w^3\, {\cal
P}(u,dw;\Delta t)$ in the same manner one can see that $\la W^3\ra
= {\cal O} (\varepsilon^2)$, we omit details. Hence we arrive at
\be
     \la W^2\ra = \frac{Q_3}{2Q_1}\, \varepsilon +
     {\cal O} (\varepsilon^2)
       \label{W20}
\ee
Substituting this into (\ref{W00}) yields
\be
      \la W\ra = \frac{Q_0Q_3}{4Q_1^2}\, \varepsilon
      + {\cal O} (\varepsilon^2)
        \label{W000}
\ee

\section{Thermodynamical analysis}
\label{secPA} \setcounter{equation}{0}

In this section we express our rigorous results in statistical
mechanical terms, such as temperature, density, pressure, and heat
transfer.

First, we consider the heat flow across the piston, from right to left.
Even though both gases are infinite, the heat transfer from one to the
other is well defined and its rate can be computed. The thermodynamical
definition of the heat transfer \cite{GF} is
\be
    R_{+\to -} = \la d{\cal E}_-/dt\ra -
    \la d{\cal M}_-/dt\ra \la W\ra
     \label{heatD}
\ee
where $d{\cal E}_-/dt$ is the rate of change of kinetic energy and
$d{\cal M}_-/dt$ the rate of change of momentum on the left hand side.
By the conservation of energy and momentum during collisions, it is
enough to compute the average change of energy and momentum of the
piston when it collides with an atom on the left hand side.

According to (\ref{V'}), the momentum of the piston at one collision
changes by
$$
     \delta {\cal M}_M = \frac{2mM}{M+m}\, (v-W)
$$
The conditional average change of momentum during the interval
$(t,t+\Delta t)$, given that $W(t)=u$, is
\begin{eqnarray*}
   E_u(\delta {\cal M}_M) &=&
   \frac{2mM}{M+m}\, \int_{D^-}(v-u)\, dx\,dv
    + {\cal O}((\Delta t)^2) \\
   &=& \frac{2mM}{M+m}\, (F_2^- - 2F_1^-u + F_0^-u^2)
   \, \Delta t + {\cal O}((\Delta t)^2)
\end{eqnarray*}
where only collisions with atoms on the left are taken into
account. Negating and averaging with respect to the stationary
measure $\mu_0$ gives the average momentum transfer to the left
gas:
\be
    \la d{\cal M}_-/dt\ra
    = -2m [ F_2^- - 2F_1^-\la W\ra + F_0^-\la W^2\ra ] +
    {\cal O}(m^2/M)
      \label{dM-}
\ee
By squaring (\ref{V'}) we find the change of the kinetic energy of the
piston at one collision:
$$
     \delta {\cal E}_M = 2m\varepsilon v^2
     + 2m Vv - 2m V^2 + \cdots
$$
where only essential terms are shown (the rest will eventually end up
in ${\cal O}(m^3/M^2)$ in the expression (\ref{dE-}) below, so we omit
them). The conditional average change of kinetic energy during the
interval $(t,t+\Delta t)$, given that $W(t)=u$, is (again only
collisions with atoms on the left are taken into account)
$$
   E_u(\delta {\cal E}_M) =
   2m (\varepsilon F_3^- -\varepsilon F_2^-u
   + F_2^-u - 2F_1^-u^2 + F_0^-u^3)
   \, \Delta t + \cdots + {\cal O}((\Delta t)^2)
$$
Negating and averaging with respect to the stationary measure
$\mu_0$ gives the average kinetic energy transfer to the left gas:
\be
    \la d{\cal E}_-/dt\ra
    = -2m [ \varepsilon F_3^- +F_2^-\la W\ra
    - 2F_1^-\la W^2\ra] +  {\cal O}(m^3/M^2)
       \label{dE-}
\ee

Combining (\ref{dM-}) and (\ref{dE-}) we obtain the expression for
the heat transfer
\be
    R_{+\to -} =
    -2m [ \varepsilon F_3^- - 2F_1^-\la W^2\ra] +  {\cal O}(\varepsilon^2)
     \label{heat}
\ee
Substituting (\ref{W20}) gives
\be
    R_{+\to -} =
    \frac{2m\varepsilon}{Q_1} (F_1^+F_3^- - F_1^-F_3^+) +  {\cal O}(\varepsilon^2)
     \label{heat0}
\ee

Now, even though the thermodynamical temperatures of the gases are not
defined, unless their velocity distributions are Maxwellian, we can
define the ``effective temperatures'' via the second moment of the
velocity distributions:
$$
    T_{\pm} = k_B^{-1}m\, \frac{\int v^2p_{\pm}\, dv}
    {\int p_{\pm}\, dv} = k_B^{-1}m\, \frac{F_2^{\pm}}{F_0^{\pm}}
$$
where $k_B$ is Boltzmann's constant. We recall that the pressures of
the gases are given by $P_{\pm}= 2mF_2^{\pm}$ and their spatial
densities by $n_{\pm}= 2F_0^{\pm}$, hence the classical law $P_{\pm}=
n_{\pm}k_BT_{\pm}$ holds.

The piston's velocity distribution $\mu_0$ is, generally, not
Maxwellian, but its effective temperature can be computed similarly:
$$
   T_M=k_B^{-1}M(\la W^2\ra - \la W\ra ^2)
   \simeq k_B^{-1}m\frac{Q_3}{2Q_1}
$$
where we used (\ref{W20}) and (\ref{W000}). Hence $T_M$ is of the same
order of magnitude as $T_{\pm}$, but for generic distributions
$p_{\pm}$ there is no simple relation between the exact values of these
temperatures.

However, we can derive some interesting relations assuming that
the distributions $p_{\pm}$ are Maxwellian. Since this would
violate our assumptions (\ref{cutoff}), our further analysis is
heuristic. Assume that the densities $p_{\pm}$ satisfy
$$
     p_{\pm}(x,v) = n_{\pm} f_{\pm}(v)
$$
where $n_{\pm}$ are (constant) spatial densities and
$$
      f_{\pm}(v) = \frac{1}{\sqrt{2\pi\sigma_{\pm}^2}}\,
      \exp\left ( -\frac{v^2}{2\sigma_{\pm}^2} \right )
$$
the Maxwellian distributions with temperatures $T_{\pm}= k_B^{-1}
m\sigma_{\pm}^2$. As before, we assume that the pressures are
equal, i.e.\ $n_-T_- = n_+T_+$. It is then easy to compute
$$
      Q_0 = \frac 12\, (n_- - n_+)
$$
$$
      Q_1 = \frac{1}{\sqrt{2\pi}}\, (n_-\sigma_- + n_+\sigma_+)
      = \frac{\sqrt{k_B}}{\sqrt{2\pi m}}\,
      \left (n_-T_-^{1/2}+n_+T_+^{1/2}\right )
$$
$$
      Q_2 = \frac 12\, (n_-\sigma_-^2 - n_+\sigma_+^2)
      = \frac{k_B}{2m}\, (n_-T_- - n_+T_+) = 0
$$
$$
      Q_3 = \sqrt{\frac{2}{\pi}}\, (n_-\sigma_-^3 + n_+\sigma_+^3)
      = \frac{\sqrt{2k_B^3}}{\sqrt{\pi m^3}}\,
      \left (n_-T_-^{3/2}+n_+T_+^{3/2}\right )
$$
Now the effective temperature of the piston is
\be
    T_M = \frac{mQ_3}{2k_BQ_1} =
    \frac{n_-T_-^{3/2}+n_+T_+^{3/2}}
    {n_-T_-^{1/2}+n_+T_+^{1/2}}
     = \sqrt{T_-T_+}
       \label{TTT}
\ee
where we used the equality of the pressures. The average velocity of
the piston is
\begin{eqnarray}
      \la W\ra &=& \frac{\sqrt{2\pi k_Bm}}{4M}\,
      \frac{(n_--n_+)(T_-T_+)^{1/2}}{n_-T_-^{1/2}+n_+T_+^{1/2}}
      + {\cal O}(\varepsilon^2) \nonumber\\
      &=& \frac{\sqrt{2\pi k_Bm}}{4M}\,\left
      (\sqrt{T_+}-\sqrt{T_-}\right)
      + {\cal O}(\varepsilon^2)
        \label{WTT}
\end{eqnarray}
Lastly, the heat transfer across the piston (from right to left) given
by (\ref{heat0}) is now
\begin{eqnarray}
    R_{+\to -} &=&
    \frac{2m\varepsilon\sqrt{2k_B^3}}{\sqrt{\pi m^3}}
    \times \frac{n_-n_+(T_+^{3/2}T_-^{1/2} - T_-^{3/2}T_+^{1/2})}
    {n_-T_-^{1/2}+n_+T_+^{1/2}} +  {\cal O}(\varepsilon^2)
    \nonumber\\
     &=&
    \frac{k_B}{M}\,\sqrt{\frac{8k_Bm}{\pi}}
    \, \frac{n_-n_+(T_-T_+)^{1/2}}
    {n_-T_-^{1/2}+n_+T_+^{1/2}}\, (T_+-T_-) +  {\cal O}(\varepsilon^2)
       \label{heatT}
\end{eqnarray}
Hence, the heat flow is proportional to the temperature gradient,
$T_+-T_-$, and the heat conductivity is thus
\be
    \kappa = \frac{k_B}{M}\,\sqrt{\frac{8k_Bm}{\pi}}
    \, \frac{n_-n_+(T_-T_+)^{1/2}}
    {n_-T_-^{1/2}+n_+T_+^{1/2}}
          \label{kappa}
\ee
The formulas (\ref{TTT})--(\ref{kappa}) were obtained earlier in
\cite{GP,GF} by means of kinetic theory.

Let us examine closely what happens when the pressures are equal,
$P_-=P_+$, but the temperatures are different. Without loss of
generality, assume that $T_-<T_+$, i.e.\ the gas on the left is cooler
but denser and the gas on the right is hotter but sparser. Then $\la
W\ra > 0$, i.e.\ the piston slowly drifts in the direction of the
hotter gas. At the same time $R_{+\to -} > 0$, i.e.\ the heat slowly
flows from the hotter gas to the cooler gas. This demonstrates that the
mechanical equilibrium is not a stable state. If our gases were finite,
there would be a continuing evolution toward {\em thermal equilibrium},
in which the temperatures become equal, too. Of course, this conclusion
makes no sense in the idealized model of infinite gases.

We note that the evolution of the system is somewhat counterintuitive.
When the piston moves to the right, i.e.\ $V(t)>0$, then by (\ref{v'})
the atoms on the right bounce off it with a higher speed and so gain
energy, while the atoms on the left collide with the piston and slow
down, hence lose some energy. When the piston moves to the left, i.e.\
$V(t)<0$, it is vice versa. Since, on the average, the heat flows from
right to left, one may conclude that the piston's movements to the left
dominate. On the other hand, the piston slowly drifts to the right, so
its displacements in that direction actually dominate.

To explain this ``paradox'', we first recall that the average speed of
the piston is of order $\la W^2\ra^{1/2}\sim \varepsilon^{1/2}$, which
is much larger than the speed of the drift $\la W\ra\sim \varepsilon$.
Hence the piston jiggles back and forth much faster than it drifts in
one direction. The heat flow is due to ``jiggling'' rather than
``drifting''. Indeed, when the piston jiggles to the right, the slow
atoms on the left have less chance to collide with the piston, since
the relative velocity is small. Most of the collisions between the slow
atoms on the left and the piston occur when the latter jiggles to the
left. This is why the cooler atoms speed up and gain energy, on the
average. On the other hand, the hotter atoms on the right are less
sensitive to the variations in the piston's velocity, and the reason
why they cool down is different, as we explain next.

The piston's vibrations are not spatially symmetric. During excursions
to the right, relatively few collisions with atoms on the right occur,
since they are more energetic and able to quickly reverse the velocity
of the piston. On the contrary, when the piston drifts to the left, the
atoms on the left are weak and it takes them longer to turn the piston
back to the right. During these intervals, the fast atoms on the right
continue hitting the piston and lose energy. This explains why the
hotter atoms mostly slow down. We must admit though that the precise
mechanism of the heat flow across the piston in our model remains
unclear. Some physicists call it a ``conspiracy'' between the
microscopic vibrations of the piston and the incoming atoms of the
gases \cite{GF,GP}.

\section{Justification of Markov approximation}
\label{secJMA} \setcounter{equation}{0}

Here we estimate the difference between the true velocity of the piston
$V(t)$ and our Markov approximation $W(t)$.

Recall that $\la W^2\ra\sim\varepsilon=m/M$, i.e. the typical speed of
the piston (in the Markov approximation) is of order
$\sqrt{\varepsilon}$. The following theorem, which is essentially
proved in \cite{CLS} (see remarks below), estimates large deviations of
the true velocity $V(t)$:

\begin{theorem}[\cite{CLS}]
Let $|V(0)|<\varepsilon^{1/2}$. There are constants $C,d>0$ such that
for every $T>0$ we have
$$
      P\left (\sup_{0<t<T}|V(t)|>
      C\, \varepsilon^{1/2}\, \ln\varepsilon^{-1}
      \right ) < T^2 \varepsilon^{-d\ln\ln\varepsilon}
$$
\label{tmVT}
\end{theorem}

This gives a good bound on large deviations during time intervals of
length $T\sim \varepsilon^{-A}$ for any large but fixed constant $A>0$:
it shows that the true velocity $V(t)$ of the piston remains ${\cal
O}(\varepsilon^{1/2}\, \ln\varepsilon^{-1})$ with ``overwhelming''
probability.

In our construction of the Markov approximation $W(t)$ we can choose
any small but positive cutoff value $\bar{V}>0$, and then assume that
$\varepsilon$ is so small that $C\, \varepsilon^{1/2}\,
\ln\varepsilon^{-1}<\bar{V}$. Then we arrive at

\begin{corollary}
For all sufficiently small $\varepsilon$ and all $T>0$ we have
$$
     P\Big (V(t) = W(t)\ \ \forall t\in (0,T)\Big )
     \geq 1 - T^2 \varepsilon^{-d\ln\ln\varepsilon}
$$
\end{corollary}

Therefore, during time intervals of length $T\sim \varepsilon^{-A}$ for
any fixed constant $A>0$, the true velocity $V(t)$ of the piston
coincides with its Markov approximation $W(t)$ with ``overwhelming''
probability. If $V(t)$ and $W(t)$ differ, though, then we have an
obvious bound: $|V(t)| \leq v_{\max}$. So we obtain analogues of our
main formulas (\ref{W000}) and (\ref{W20}):

\begin{corollary}
Let $\varepsilon>0$ be small and $|V(0)|<\varepsilon^{1/2}$. Then for
any fixed $A>0$ and all $0<t<T=\varepsilon^{-A}$ we have
$$
      E( V(t) ) = \frac{Q_0Q_3}{4Q_1^2}\, \varepsilon
      + {\cal O} (\varepsilon^2)
$$
and
$$
     E( V^2(t) ) = \frac{Q_3}{2Q_1}\, \varepsilon +
     {\cal O} (\varepsilon^2)
$$
where $E(\cdot )$ is the mean value. \label{crEVVV}
\end{corollary}

It follows from the results of \cite{CLS} that whenever the true
velocity $V(t)$ exceeds $C\, \varepsilon^{1/2}\, \ln\varepsilon^{-1}$,
then with ``overwhelming'' probability it will be driven back toward
zero by further collisions with the gas atoms. Hence, the estimates of
the last corollary should remain true for all $t>0$, but we do not
pursue such a goal, because the drift of the piston becomes evident on
the time scale of order $\varepsilon^{-A}$ already for $A>2$, as we
show next, so the estimates in Corollary~\ref{crEVVV} are sufficient
for practical purposes.

Indeed, let $X(0)=0$ and consider the random position of the piston
$X(t)$ at time $t$. Also, let $Y(t)=\int_0^t W(s)\, ds$ be the
corresponding function in the Markov approximation model. We obviously
have
\be
        \la Y(t)\ra = t\la W\ra =\frac{Q_0Q_3}{4Q_1^2}\, \varepsilon t
           + {\cal O} (\varepsilon^2t)
             \label{EY1}
\ee
We also need to estimate the variance of $Y(t)$, which is a little
harder. We claim that
\be
      {\rm Var}[Y(t)] \leq D t
         \label{EY2}
\ee
with some constant $D>0$ independent of $\varepsilon$. Below we outline
the argument, suppressing some technical details that are based on the
estimates obtained in \cite{CLS}.

It is standard that
$$
    {\rm Var}[Y(t)] = \int_0^t\int_0^t{\rm Cov}(W(s),W(u))\, ds\, du
$$
By the stationarity of the process $W(t)$, the covariance here equals
$$
     {\rm Cov}(W(s),W(u))=\rho_{s-u}\, {\rm Var}(W)
$$
where $\rho_{s-u}$ is the correlation between $W(s)$ and $W(u)$, which
only depends on $|s-u|$. Also recall that
$$
      {\rm Var}(W) = \la W^2\ra - \la W\ra^2 =
      \frac{Q_3}{2Q_1}\, \varepsilon + {\cal O} (\varepsilon^2)
$$
Therefore,
\be
           {\rm Var}[Y(t)] \leq {\rm const}\cdot t\, \varepsilon\,
           \int_0^t  \rho_s\, ds
              \label{VarY}
\ee
It remains to estimate the decay of the velocity autocorrelation
function $\rho_s$.

Suppose during the interval $(u,u+s)$ the piston experiences $k$
collisions with gas atoms with velocities $v_1,\ldots,v_k$ numbered in
the order in which the collisions occur. Then (\ref{V'}) implies
\be
    W(u+s)=(1-\alpha)^k W(u) +
    \alpha \sum_{j=1}^k (1-\alpha)^{k-j} \cdot v_j
       \label{main}
\ee
where $\alpha=2m/(M+m)\simeq 2\varepsilon$. This equation can be easily
verified by induction on $k$.

Now, the number of collisions $k$ grows as const$\cdot s$, on the
average, in fact
\be
      P(|k - c_1s| \geq c_2\sqrt{s}\ln s) \leq
      s^2 \varepsilon^{-d\ln\ln\varepsilon}
        \label{ks}
\ee
with some positive constants $c_1,c_2$, see \cite{CLS}. For
$s<\varepsilon^{-A}$ this gives a good bound on large deviations, so we
can safely assume that $k\geq c_1\varepsilon/2$. Next, the velocities
$v_i$ are almost independent of $W(u)$. In fact, there is a dependence,
since the time of collision between a particular atom and the piston
depends on the velocity of the latter, but the correlation between
$W(u)$ and velocities $v_j$ for every fixed $j$ is negative. This
follows from a simple observation: if $W(u)>0$, the piston is more
likely to collide with atoms on the right, for which $v_j<0$, and vice
versa. Therefore, the correlation $\rho_s$ is determined by the first
term in (\ref{main}), hence
$$
       |\rho_s| \leq (1-\alpha)^{c_1s/2}
$$
This implies
$$
      \int_0^t \rho_s\, ds \leq
      \int_0^{\infty}(1-\alpha)^{c_1s/2}\,ds
      = -\frac{2}{c_1\ln (1-\alpha)} \simeq
      \frac{1}{c_1\varepsilon}
$$
and combining with (\ref{VarY}) proves (\ref{EY2}).

The estimates (\ref{EY1}) and (\ref{EY2}) show that the piston's
displacement due to the drift exceeds its typical random fluctuations
at times $t=\varepsilon^A$ for all $A>2$. On such time scales the drift
is ``physically evident'' and it is possible to verify it
experimentally. Numerical experiments of this sort were done by
\cite{GF}.

We now comment on the proofs of Theorem~\ref{tmVT} and the estimate
(\ref{ks}). The corresponding results were proven in \cite{CLS} in a
slightly different context. That paper dealt with a heavy piston in a
cubical container of side $L$ filled by an ideal gas. That model, too,
reduced by a trivial projection onto the axis perpendicular to the
piston surface to a one-dimensional gas on an interval $[-L/2,L/2]$.
The velocity cutoff assumptions in \cite{CLS} were identical to our
(\ref{cutoff}). The piston had mass $M=aL^2$ ($a>0$ was a constant) and
the spatial density of the one-dimensional gases was proportional to
$M=aL^2$, but a simple rescaling of time and space by $M$ reduces the
spatial density to values ${\cal O}(1)$ and makes it independent of
$M$. Hence, we arrive exactly at the same initial conditions as in the
present paper, except the gases in \cite{CLS} were finite -- they
evolved on an interval $[-aL^3/2,aL^3/2]$ with the piston initially
placed at $X(0)=0$.

Due to the velocity cutoff (\ref{cutoff}) the gas atoms in \cite{CLS}
cannot interact with the piston more than once unless they travel
across the half interval $[-aL^3/2,aL^3/2]$, reflect at an endpoint
$x=\pm aL^3/2$ and then travel back to the piston in the middle. This
takes time $T={\cal O}(L^3)$, as it was proven in \cite{CLS}. Thus,
during the initial interval $(0,T)$ with $T={\cal O}(M^{3/2})$ the
evolution of the piston studied in \cite{CLS} is identical to ours.
Hence, all the results and estimates derived in Section~3 of \cite{CLS}
directly apply to our present context. Furthermore, with some minor
obvious changes those results can be extended to arbitrary time
intervals as we did in Theorem~\ref{tmVT} and (\ref{ks}).

Lastly, we have assumed the velocity cutoff (\ref{cutoff}) here and in
the paper \cite{CLS} mostly for convenience. It might be true that the
upper bound on velocities is quite essential in \cite{CLS}, since
arbitrarily fast atoms could bounce back and forth between the piston
and the wall many times during short time intervals thus making
significant impact on the piston, as it was remarked in Section~5 of
\cite{CLS}. But in the present work, if we had arbitrarily fast
particles, they would hit the piston once and get away never to come
back, so their impact would be quite limited.

As for the lower bound on velocities, it seems to be less essential in
both models. It was conjectured in Section~5 of \cite{CLS} that the
results of that paper could be extended to velocity distributions
without lower bounds (i.e.\ with $v_{\min}=0$). If so, then the
estimate in Theorem~\ref{tmVT} implies that the typical velocity of the
piston remains of order $\varepsilon^{1/2}\, \ln\varepsilon^{-1}$. As
long as this is true, only atoms with smaller velocities can experience
two or more collisions with the piston, thus violating the Markovian
character of the piston's velocity process $V(t)$. It is customary to
redefine the dynamics so that the collisions with slow atoms are
ignored altogether, and then the resulting process $W(t)$ will be
Markovian. The difference $|V(t)-W(t)|$ should be estimated separately
and presumably is negligible. Such a strategy was successfully
implemented in earlier works \cite{H,DGL} in somewhat different
context, and it is likely to go through in our case as well. This is
the subject of our future work.

\medskip\noindent
{\bf Acknowledgement}. The author thanks J.~Lebowitz, J.~Piasecki
and Ya.~Sinai for many useful discussions and encouragement. The
author was partially supported by NSF grant DMS-0098788. This work
was completed during the author's stay at the Institute for
Advanced Study with partial support by NSF grant DMS-9729992.

\end{document}